\documentclass[12pt]{article}

\def\(#1){(\ref{#1})}           

\def\sft{{square Fibonacci tiling}}

\def\kv{{\bf k}}

\def\rv{{\bf r}}

\def\={{\equiv}}

\begin{document}
\thispagestyle{empty}

\begin{LARGE} 
\begin{center}
%
%
Quasicrystals: A matter of definition.
%
%
\end{center}
\end{LARGE}

\begin{large}
\begin{center} 
%
%
Ron Lifshitz
%
%
\end{center}
\end{large}

%
%
\begin{it}
\begin{center}
School of Physics \& Astronomy, The Raymond and Beverly Sackler
Faculty of Exact Sciences, Tel Aviv University, Tel Aviv 69978, Israel
\end{center}
\end{it}
%
%

\begin{center}
%
%
\bigskip\bigskip
%
%
\end{center}

%
%
It is argued that the prevailing definition of quasicrystals,
requiring them to contain an axis of symmetry that is forbidden in
periodic crystals, is inadequate. This definition is too restrictive
in that it excludes an important and interesting collection of
structures that exhibit all the well-known properties of quasicrystals
without possessing any forbidden symmetries.
%
%
\vspace{6ex}

%
%
\vspace{6ex}

%
%

\section{Introduction}

The aim of this paper is to argue against the common practice$^{1-4}$
to restrict the definition of quasicrystals by requiring that they
possess an axis of symmetry that is incompatible with periodicity.
According to this restriction there are no quasicrystals in
1-dimension, and a quasicrystal in 2- or 3-dimensions must have an
axis of $N$-fold symmetry, with $N=5$, or $N>6$. I propose here to
accept the original definition of Levine and Steinhardt$^5$ whereby
the term {\it quasicrystal\/} is simply an abbreviation for {\it
  quasiperiodic crystal,} possibly with the proviso that the term
quasicrystal be used for crystals that are strictly aperiodic (as the
mathematical definition of quasiperiodicity includes periodicity as a
special case).

I shall start by reviewing some basic definitions in
section~\ref{sec:def}. I shall then proceed in section~\ref{sec:fam}
to discuss the problematic distinction between the different families
of quasiperiodic crystals, namely, incommensurately modulated
crystals, incommensurate composite crystals, and those crystals that
are typically referred to as quasicrystals.  Finally, in
section~\ref{sec:examples}, I shall support my call to relax the
definition of quasicrystals by referring to theoretical models, as
well as experimental observations, of structures which should be
considered as quasicrystals even though they possess no forbidden
symmetries.

\section{Definitions}
\label{sec:def}

\subsection{What is a crystal?}

Before Shechtman's 1982 discovery of the first quasicrystal$^6$ it was
universally accepted, though never proven, that the internal order of
crystals was achieved through a periodic filling of space.
Crystallography treated {\it order\/} and {\it periodicity\/}
synonymously, both serving equally to define the notion of a {\it
  crystal.} With that came the so-called ``crystallographic
restriction,'' stating that crystals cannot have certain forbidden
symmetries, such as 5-fold rotations.  The periodic nature of crystals
was ``confirmed'' with the discovery of x-ray crystallography and
numerous other experimental techniques throughout the 20th century.
Periodicity became the underlying paradigm, not only for
crystallography itself, but also for other disciplines such as
materials science and condensed matter physics, whose most basic
tools, like the Brillouin zone, rely on its existence.

Two decades later, it is now clear that periodicity and order are not
synonymous, and that a decision has to be made as to which should
define the term {\it crystal.} The International Union of
Crystallography through its Commission on Aperiodic Crystals$^7$
decided on the latter, but was not ready to give precise microscopic
descriptions of all the ways in which order can be achieved. Clearly,
periodicity is one way of achieving order, quasiperiodicity as in
Penrose-like tilings is another, but can we be certain that there are
no other ways that have not yet been discovered?  The Commission opted
to shift the definition from a microscopic description of the crystal
to a property of the data collected in a diffraction experiment. It
decided on a temporary working-definition whereby a {\it crystal\/} is
``any solid having an essentially discrete diffraction diagram.''
Crystals that are periodic are explicitly called {\it periodic
  crystals,} all others are called {\it aperiodic crystals.} The new
definition is consistent with the notion of long-range order, used in
condensed matter physics, where the transition from a disordered
liquid to an ordered solid is indicated by the appearance of an {\it
  order parameter\/} in the form of Bragg peaks in the diffraction
diagram at non-zero wave vectors. It is sufficiently vague so as not
to impose unnecessary constraints until a better understanding of
crystallinity emerges. We need not worry about this vagueness here,
because we shall only be concerned with quasiperiodic crystals, which
are a well defined subcategory of structures, satisfying the new
definition.

\subsection{What is a quasiperiodic crystal?}

Solids whose density functions $\rho(\rv)$ may be expanded as a
superposition of a countable number of plane waves 
\begin{equation}
  \label{eq:fourier}
  \rho(\rv) = \sum_{\kv\in L}\rho(\kv) e^{i\kv\cdot\rv},
\end{equation}
are called {\it almost periodic crystals}.  In particular, if taking
integral linear combinations of a finite number $D$ of wave vectors in
this expansion can span all the rest, then the crystal is {\it
  quasiperiodic.} The diffraction pattern of a quasiperiodic crystal,
therefore, contains Bragg peaks each of which can be indexed by $D$
integers.  If $D$ is the smallest number of wave vectors that can span
the whole set $L$ using integral linear combinations then $D$ is
called the {\it rank,} or the {\it indexing dimension\/} of the
crystals. {\it Periodic crystals\/} form a special subset of all
quasiperiodic crystals whose rank $D$ is equal to the actual physical
dimension $d$.  For periodic crystals the set of Bragg peaks is truly
discrete because the set of wave vectors $\kv$ in their Fourier
expansion (\ref{eq:fourier}) is discrete. For quasiperiodic crystals
whose rank $D$ is greater than the physical dimension $d$, the set $L$
of wave vectors in the expansion (\ref{eq:fourier}) is dense---there
are $\kv$'s in $L$ that cannot be surrounded by a finite
$d$-dimensional ball that contains no other $\kv$'s. Nevertheless, in
actual experiments, where the total integrated diffraction intensity
is finite, Bragg peaks are not observed at wave vectors $\kv$ for
which the intensity $|\rho(\kv)|^2$ is below a certain threshold. The
observed diffraction pattern is therefore essentially discrete even
though the set $L$ is not. It should be noted that all experimentally
observed crystals to date are quasiperiodic.

\section{The quasicrystallographic restriction}
\label{sec:fam}

Certain classes of quasiperiodic crystals were known long before
Shechtman's discovery. These are the so-called {\it
  incommensurately-modulated crystals\/} and {\it incommensurate
  composite crystals,} (or {\it intergrowth compounds\/}). The former
consist of a basic (or average) ordered structure that is perturbed
periodically (modulated) in space, and the period of the modulation is
incommensurate with the underlying spatial periodicities of the basic
structure.\footnote{We know today of cases where the basic structure
  itself is already aperiodic.$^{8,9}$} The latter are composed of two
or more interpenetrating subsystems with mutually incommensurate
spatial periodicities.  Each subsystem, when viewed independently, is
itself a crystal which is incommensurately-modulated due to its
interaction with the other subsystems. The diffraction diagrams of
these special types of quasiperiodic crystals are characterized by
having one or more subsets of {\it main reflections}---Bragg peaks
that are significantly brighter than the others---describing the basic
structures, and weaker peaks, called {\it satellites,} arising from
the modulations. For more detail see, for example, references 2 and
10.

Incommensurately-modulated and incommensurate composite crystals did
not pose any serious challenge to the periodicity paradigm because
they could all be viewed as periodic structures that had been slightly
modified.  Order was still obtained through periodicity---the paradigm
remained intact. Shechtman's discovery implied that there exist
quasiperiodic crystals for which a description in terms of a
modulation of a basic periodic structure or a composition of two or
more substructures is either inappropriate or impossible. Due to its
forbidden 5-fold symmetry, Shechtman's quasicrystal was clearly
not a quasiperiodic modification of a periodic crystal, but rather a
crystal which was somehow intrinsically quasiperiodic---a crystal in
which order was {\it not\/} achieved by means of periodicity.
Shechtman's discovery was able to shatter the old paradigm because it
was a clear violation of the crystallographic restriction. The
observation of a forbidden symmetry was so pivotal in the discovery of
quasicrystals that it became their defining property. The
crystallographic restriction was replaced by what may be viewed as a
``quasicrystallographic restriction.''

It is common practice to reserve the term ``quasicrystal'' exclusively
for those crystals, like Shechtman's, that are intrinsically
quasiperiodic, setting them apart from modulated and composite
crystals as a third subcategory of quasiperiodic structures. This
common point of view\footnote{See, for example, references 2, 10, 11,
  and 12.} is appealing for many reasons, particularly, because there
are systems whose physical behavior is indeed governed by the fact
that the crystal is modulated or composed of substructures. Not
viewing these systems as such, and not utilizing the many theoretical
and experimental tools developed specifically for treating modulated
and composite crystals, would be foolish.

The problem with the desire to distinguish between
intrinsically-quasiperiodic crystals and crystals in which
quasiperiodicity is obtained via modulation or composition is the lack
of a quantitative criterion for making this distinction. The easiest
way to see the difficulty is by considering the diffraction patterns.
The diffraction pattern of a modulated crystal, for example, must
exhibit a subset of strong main reflections accompanied by weak
satellites.  This begs to ask how weak must the satellites be to be
considered as such? If one could hypothetically gradually increase the
intensity of the satellites and their harmonics, at what point would
the structure cease to be a modulated crystal? The same difficulty can
also be seen in direct space, by starting with a periodic crystal
which is modulated by a smooth incommensurate sine function, and
gradually increasing the amplitude of this modulation while adding
higher harmonic contributions. If consequently the modulation takes
the shape of an unsmooth sawtooth function would it not be more
appropriate to view it as a set of separate ``atomic surfaces'' like
one does in a quasicrystal? 

It turns out that this gradual transformation of a modulated crystal
into a ``quasicrystal'' is not at all hypothetical.  There are
examples of systems,$^{13}$ in which this transformation happens as a
function of composition. The transformed structures are described as
modulated crystals, with complicated modulation functions, called
``Crenel functions'',$^{14}$ when in fact they can be described very
simply as ``quasicrystals'' with simple atomic surfaces, as explained
in Ref.~13.\footnote{This may remind the reader of the famous
  experiment in which a group of people is shown a sequence of
  pictures, beginning with a cat which gradually changes into a dog.
  The viewers insist that they are still seeing a cat almost to the
  end, when in fact they looking at a picture of a dog.}  Although,
one of the best suggestions$^{12}$ for the distinction between
quasicrystals and modulated crystals is based on the shape of the
modulation function, it seems quite impractical as a quantitative
experimental criterion.

So, even though there are clearly structures that are formed by
modulating or composing simpler structures, and there are clearly
other structures that are not, there is simply no quantitative
criterion to distinguish between these categories of quasiperiodic
structures. Unless of course, as a last resort, one adopts the
quasicrystallographic restriction. The criterion is then very simple:
If a quasiperiodic crystal possesses forbidden symmetries then it
is a quasicrystal, otherwise it is a modulated or a composite crystal.
This is probably the most appealing reason to adopt the
quasicrystallographic restriction. The problem is that it leaves no
room for the possible existence of crystals that are intrinsically
quasiperiodic---not formed by modulation or composition---yet possess
no forbidden symmetries. If such crystals cannot exist then there
is no problem with adopting the quasicrystallographic restriction. If
such crystals do exist then adopting the restriction would be
inappropriate. So we must ask: Are there are any examples of such
crystals?

\section{Many examples}
\label{sec:examples}

Theoretical models of such crystals, that are intrinsically
quasiperiodic yet possess no forbidden symmetries, are very easy to
construct. In fact, from a theoretical standpoint it should be {\it
  obvious\/} that there is nothing special about point groups that are
incompatible with periodicity. In principle, any method that is used to
generate a quasiperiodic tiling with, say, 10-fold symmetry can be
used to generate quasiperiodic tilings with, say, 4-fold symmetry.
Indeed, there are many examples in the literature of tiling models of
quasicrystals, with 2-, 4-, and 6-fold symmetry, generated by all the
standard methods: matching rules,$^{15}$ substitution rules,$^{15,16}$
the cut-and-project method$^{17,18}$ and the dual-grid method.$^{19}$

I have recently described the two-dimensional \sft\ and its natural
generalization into three (or even higher) dimensions.$^{20}$ It is a
quasiperiodic tiling with many of the features normally associated
with standard tiling models of quasicrystals like the Penrose tiling.
It has a finite number of tiles with definite tile frequencies and a
finite number of vertex configurations; it can be generated by most of
the standard methods for generating quasiperiodic tilings; its
diffraction diagram contains Bragg peaks with no clear subset of
main-reflections; and most notably, it has $\tau$-inflation symmetry,
where $\tau=(1+\sqrt{5})/2$ is the golden ratio.  Like the proverbial
bird that looks like a duck, walks like a duck, quacks like a duck,
and is therefore a duck---the \sft\ {\it is\/} a model quasicrystal
even though it has no forbidden symmetries.

To the best of my knowledge, no alloys or real quasicrystals exist
with the precise structure of the square or cubic Fibonacci tilings.
Yet, this does not imply that structures like the \sft\ are
experimentally irrelevant. In recent years we have come to know a
number of experimental applications where one creates artificial
quasicrystals.  One example is in the field of photonic
crystals,$^{21}$ with the aim of producing novel photonic band-gap
materials. Another example is in field of non-linear optics,$^{22}$
with the aim of achieving third- and fourth-harmonic generation in a
single crystal. In both of these examples it may be beneficial to make
artificial quasicrystals with structures, similar to that of the \sft.

The existence of theoretical models and the possibility to fabricate
artificial structures might be dismissed as trivial, yet the existence
of actual experimental observations is a different matter. It turns
out that there have been experimental reports of quasiperiodic
crystals with cubic symmetry$^{23,24}$ as well as
tetrahedral,$^{25,26}$ tetragonal,$^{17}$ and possibly also
hexagonal$^{27}$ symmetry, that are neither modulated crystals nor
composite crystals. Their diffraction diagrams show no clear subset(s)
of main reflections, yet they do not possess any forbidden symmetry.
One of the cubic quasicrystals,$^{24}$ a Mg-Al alloy, is even reported
to have inflation symmetry involving irrational factors related to
$\sqrt{3}$.  These crystals are clearly quasiperiodic yet they are not
formed by modifying an underlying periodic structure. They are as
intrinsically quasiperiodic as the quasicrystals that have forbidden
symmetries, and should therefore all be considered quasicrystals.

\section{So, what is a quasicrystal?}

I suggest that the quasicrystallographic restriction, requiring
quasicrystals to possess forbidden symmetries, be officially
abandoned.  I would like the scientific community to accept the
original definition of Levine and Steinhardt$^5$ whereby the term {\it
  quasicrystal\/} is simply an abbreviation for {\it quasiperiodic
  crystal,} possibly with the proviso that the term quasicrystal be
used only for crystals that are strictly aperiodic (since, as mentioned
above, the mathematical definition of quasiperiodicity includes
periodicity as a special case).

This paper is part of an ongoing debate on the meaning of
crystallinity and the concept of a quasicrystal. Some
crystallographers might still be under the impression that if a
quasiperiodic crystal does not possess any forbidden symmetry it must
be either an incommensurately modulated crystal or an incommensurate
composite crystal, and that no other possibility exists. Many
crystallographers still impose the ``quasicrystallographic
restriction'' when defining quasicrystals in their publications. It is
my firm opinion that these practices and misconceptions should be
stopped, not only as a matter of academic preciseness, but more
importantly, to make sure that crystallographers who discover new
quasicrystals without forbidden symmetries will not hesitate to
publish their findings. As we celebrate in these Journal issues the
many contributions of David Mermin to science, its teaching,$^{28}$
and its communication to others,$^{29}$ a more appropriate title for
this article (in the spirit of David Mermin's ``Reference Frame''
columns in {\it Physics Today\/}) might have been ``What's wrong with
these quasicrystals?'' The answer in this case is that nothing is
wrong with {\it these\/} quasicrystals---the problem lies with the
definition.

This paper is dedicated to David Mermin on the occasion of his first
steps towards retirement. I would like to take this opportunity to
thank David once again for being such a great teacher and a
wonderful collaborator.  

This research is supported by the Israel Science Foundation through
Grant No.~278/00.

%
%

%
%

\section*{References}

\vspace{3ex}
\itemsep 0.2ex
\parindent=0pt

\begin{itemize}
\item[1.] P.~M.~Chaikin and T.~C.~Lubensky, {\it Principles of Condensed Matter
   Physics} (Cambridge University Press, Cambridge, 1995), p. 680.

\item[2.] S.~Van Smaalen, ``Incommensurate crystal structures,'' {\it
    Cryst.\ Rev.} {\bf 4}, 79 (1995). 
  
\item[3.] M.~Senechal, {\it Quasicrystals and Geometry} (Cambridge
  University press, Cambridge, 1996), p. 31 (but see the comment on p. 33).

\item[4.] D.~Gratias, ``A brief survey of quasicrystallography,''
  {\it Ferroelectrics\/} {\bf 250} 1 (2001).
  
\item[5.] D.~Levine and P.~J.~Steinhardt, ``Quasicrystals I:
  Definition and structure,'' {\it Phys. Rev.} B{\bf 34},
  596 (1986).
  
\item[6.] D.~Shechtman, I.~Blech, D.~Gratias, and J.~W.~Cahn,
   ``Metallic phase with long-ranged orientational order and no
   translational symmetry,'' 
   {\it Phys. Rev. Lett.} {\bf 53}, 1951 (1984).
   
\item[7.] International Union of Crystallography, {\it Acta Cryst.}
   A{\bf 48}, 922 (1992).

\item[8.] F.~Denoyer, P.~Launois, T.~Motsch, and M.~Lambert, ``On the
  phase transition mechanism in Al-Cu-Fe: structural analysis of the
  modulated quasicrystalline and of the microcrystalline states,'' 
  {\it J.\ Non-Cryst.\ Solids\/} {\bf 153 \& 154}, 595 (1993).

\item[9.] N.~Menguy, M.~de~Boissieu, P.~Guyot, M.~Audier, E.~Elkaim,
  and J.~P.~Lauriat, ``Modulated icosahedral Al-Fe-Cu phase: a single
  crystal X-ray diffraction study,'' {\it Ibid.} p. 620.
  
\item[10.] R.~Lifshitz, ``The symmetry of quasiperiodic crystals,''
  {\it Physica\/} A{\bf 232}, 633 (1996).  
  
\item[11.] T.~Janssen, ``Crystallography of quasicrystals,'' {\it Acta
    Cryst.} A{\bf 42}, 261 (1986).

\item[12.] A.~Katz and D.~Gratias, in {\it Lectures on Quasicrystals,}
   ed. F. Hippert and D. Gratias (Les Editions de Physique, Les Ulis,
   1994), see p. 247.

\item[13.] L.~Elcoro, J.~M.~Perez-Mato, and R.~Withers,
  ``Intergrowth polytypoids as modulated structures: The 
  example of the cation deficient oxides LaTi1-xO3,'' {\it Zeit.\
    Krist.} {\bf 215}, 727 (2000). 
  
\item[14.] V.~Pet\v{r}\'\i\v{c}ek, A.~van der Lee and M.~Evain, ``On
  the use of Crenel functions for occupationally modulated
  structures,'' {\it Acta Cryst.} A{\bf 51}, 529 (1995).

\item[15.] B.~Gr\"unbaum and G.~C.~Shephard, {\it Tilings and Patterns}
    (Freeman, New York, 1987), section 10.4.
    
\item[16.] R.~Paredes, J.~L.~Arag\'on, and R.~A.~Barrio, ``Nonperiodic
  hexagonal square-triangle tilings,'' {\it Phys.\ Rev.} B{\bf 58},
  11990 (1998). 
    
\item[17.] Z.~H.~Mai, L.~Xu, N.~Wang, K.~H.~Kuo, Z.~C.~Jin, and
  G.~Cheng, ``Effect of phason strain on the transition of an
  octagonal quasicrystal to a $\beta$-Mn-type structure,'' {\it
    Phys. Rev.} B{\bf 40}, 12183 (1989). 
  
\item[18.] M.~Baake, D.~Joseph, P.~Kramer, ``The Schur rotation as a
  simple approach to the transition between quasiperiodic and periodic
  phases,'' {\it J.\ Phys.\ A: Math.\ Gen.} {\bf 24}, L961 (1991).
    
\item[19.] S.~Baranidharan and E.~S.~R.~Gopal, ``Quasiperiodic tilings
  with fourfold symmetry,'' {\it Acta Cryst.} A{\bf 48}, 782 (1992).

\item[20.] R.~Lifshitz, ``The square Fibonacci tiling,'' {\it J.
    Alloys Comp.} {\bf 342}, 186 (2002). 

\item[21.] M.~Hase, M.~Egashira, N.~Shinya, H.~Miyazaki, K.M.~Kojima,
  S.~Uchida,  ``Optical transmission spectra of two-dimensional
  quasiperiodic photonic crystals based on Penrose-tiling and
  octagonal-tiling systems,'' {\it J. Alloys Comp.} {\bf 342}, 455 (2002).
  
\item[22.] K.~Fradkin-Kashi, A.~Arie, ``Multiple-wavelength
  quasi-phase-matched nonlinear interactions,'' {\it IEEE J. Quantum
    Electron.}  {\bf 35}, 1649 (1999); K.~Fradkin-Kashi, A.~Arie,
  P.~Urenski, G.~Rosenman, ``Multiple nonlinear optical interactions
  with arbitrary wave vector differences,'' {\it Phys. Rev. Lett.}
  {\bf 88}, 023903 (2002).

\item[23.] Y.~C.~Feng, G.~Lu, H.~Q.~Ye, K.~H.~Kuo, R.~L.~Withers, and
  G.~Van Tendeloo, ``Experimental evidence for and a projection model
  of a cubic quasicrystal,'' {\it J.\ Phys.: Condens.\ Matter\/} {\bf
    2}, 9749 (1990). 

\item[24.] P.~Donnadieu, H.~L.~Su, A.~Proult, M.~Harmelin,
  G.~Effenberg, and F.~Aldinger, ``From modulated phases to a
  quasiperiodic structure with a cubic point group and inflation
  symmetry,'' {\it J. Phys. I France\/} {\bf 6}, 1153 (1996).

\item[25.] P.~Donnadieu, ``The deviation of the Al6Li3Cu quasicrystal
  from icosahedral symmetry - A reminisence of a cubic crystal,'' {\it
    J. Phys. I France\/} {\bf 4}, 791 (1994). 
  
\item[26.] J.~Dr\"ager, R.~Lifshitz, and N.~D.~Mermin, ``Tetrahedral
  quasicrystals,'' in
  {\it Proceedings of the 5th International Conference on
    Quasicrystals,} ed. C. Janot and R. Mosseri (World Scientific,
  Singapore, 1996) 72.
  
\item[27.] H.~Selke, U.~Vogg, and P.~L.~Ryder, ``New quasiperiodic
  phase in Al85Cr15,'' {\it Phys.\ Stat.\ Sol.} A{\bf 141}, 31 (1994).

\item[28.] N.~W.~Ashcroft and N.~D.~Mermin, {\it Solid state physics}
  (Saunders College, Philadelphia, 1976). 

\item[29.] N.~D.~Mermin, {\it Boojums all the way through} (Cambridge
  University Press, Cambridge, 1990).

\end{itemize}

\end{document}